\documentclass[a4paper]{jpconf}
\usepackage{graphicx}
\begin{document}
\title{Weakly bound systems, continuum effects, and reactions}

\author{Y Jaganathen$^1$, N Michel$^{2,3}$ and M P{\l}oszajczak$^1$}

\address{$^1$ Grand Acc\'el\'erateur National d'Ions Lourds (GANIL), CEA/DSM - CNRS/IN2P3, BP 55027, F-14076 Caen Cedex, France} 
\address{$^2$ Department of Physics and Astronomy, University of Tennessee, Knoxville, Tennessee 37996, USA}
\address{$^3$ Physics Division, Oak Ridge National Laboratory, Oak Ridge, Tennessee 37831, USA}

\ead{ploszajczak@ganil.fr}

\begin{abstract}
Structure of weakly bound/unbound nuclei close to particle drip lines is different from that around the valley of beta stability. A comprehensive description of these systems  goes beyond standard Shell Model (SM) and demands an open quantum system description of the nuclear many-body system. We approach this problem using the Gamow Shell Model (GSM) which provides a fully microscopic description of bound and unbound nuclear states, nuclear decays, and reactions. We present in this paper the first application of the GSM for a description of the elastic and inelastic scattering of protons on $^6$He.

\end{abstract}

\section{Introduction}
The nuclear SM was proposed almost sixty years ago\cite{Gop49,Hax49}. Soon after, the interacting SM was developed and used extensively to understand a wealth of data on nuclear levels,  moments, collective excitations, electromagnetic and  $\beta$ decays, and various particle decays. Interacting SM describes the nucleus as a closed quantum system: nucleons occupy bound, hence well localized,  single-particle orbits of a harmonic oscillator potential and are isolated from the environment of unbound scattering states. Since the scattering continuum is not considered explicitly, the presence of decay thresholds and the interplay of Hermitian (internal) and anti-Hermitian (external, via the continuum) mixing of SM configurations is neglected. This competition yields a complicate interference pattern\cite{rf:28} and is a source of many collective features such as, e.g., the resonance trapping\cite{rf:12,rf:13,rf:14} and super-radiance phenomenon\cite{rf:15,rf:16},  the multichannel coupling effects in reaction cross-sections\cite{rf:19} and shell occupancies\cite{rf:20},  
the clustering\cite{rf:40}, the modification of spectral fluctuations\cite{rf:17}, the deviations from Porter-Thomas resonance widths distribution\cite{rf:14,rf:18}, and so on. 

It was clear in fifties that the SM is a phenomenological tool which employs the strongly renormalized bare nuclear interaction between nucleons\cite{Brueckner} and neglects coupling to the continuum\cite{Wigner}. The overwhelming success of the SM, its elegance and simplicity resulted in neglecting most of these questionable assumptions.  Unfortunately, the divide between discrete and continuum states in SM has led to an artificial separation of nuclear structure from nuclear reactions, and hindered a deeper understanding of nuclear properties. It is often believed that the understanding why the simple interacting SM works so well will be advanced when the goal of {\em ab initio} many-body approaches including the continuum coupling will be achieved. 

Many structural properties of the nucleus are determined by means of nuclear collisions and this calls for a unified theoretical framework. Feshbach at the beginning of sixties formulated a unified theory of nuclear reactions using the effective Hamiltonian and the projection operator method to select the open channel components of the wave function\cite{fesh}. This development led to various formulations of the real-energy continuum shell model (CSM)\cite{Mah69,Bar77,Phi77,Ben99,Vol06,Rot06}. Nowadays, the real-energy CSM provides a unified description of the structure and reactions with up to two nucleons in the scattering continuum\cite{Ben99,Rot06}. 

At the same time, Fano noticed that the exact coincidence of different configurations above the lowest particle emission threshold makes the perturbation theory inadequate and calls for a generalization of the standard SM\cite{fano}. The achievement of Fano's goal took almost forty years and required the development of new mathematical approach of the rigged Hilbert space\cite{gel}, new methods to deal with diverging integrals (matrix elements of one- and two-body operators) involving resonance and scattering states\cite{Zel60,Hok65,Rom68,Zim70,Gya71}, the formulation of the generalized completeness relation including single-particle (s.p.) bound states, resonances and scattering states\cite{berggren}. These different and independent developments in mathematics and physics enabled finally a satisfactory formulation of the new many-body theory, the GSM\cite{Mic02,Idb02,Mic03,Mic04}, which offers a fully symmetric treatment of bound, resonance and scattering states in the multiparticle framework. In this formulation, the maximum number of particles in the scattering continuum is not  {\it a priori} prescribed as in the real-energy CSM, but follows from the Schr\"{o}dinger variational principle for the many-body Hamiltonian.

Until now, GSM has been primarily used  in the context of nuclear structure.  (For a recent review, see Ref. \cite{Mic09}.)
In this paper, we shall extend GSM  to reaction problems by the  coupled-channel (CC) formulation of the scattering problem. The very first application of the GSM-CC formalism will be presented in this paper for the proton scattering on $^6$He target. The proposed GSM-CC formalism can be easily straightforwardly generalized for the description of nuclear reactions in the {\it ab initio} framework of the No-Core Gamow Shell Model\cite{papa12}.

In Sect. 2 we present essential features of the GSM. The reaction wave function and the derivation of GSM-CC equations in the coordinate space representation are discussed in Sect. 3. The method for solving the CC equations is presented in Sect. 4. The first application of the GSM-CC formalism for a description of $^6$He+p reaction is discussed in Sect. 5. Finally, main conclusions of this work are summarized in Sect. 6.

\section{The Gamow Shell Model }
\begin{figure*}
\begin{center}
 \includegraphics[width=11cm,angle=-90]{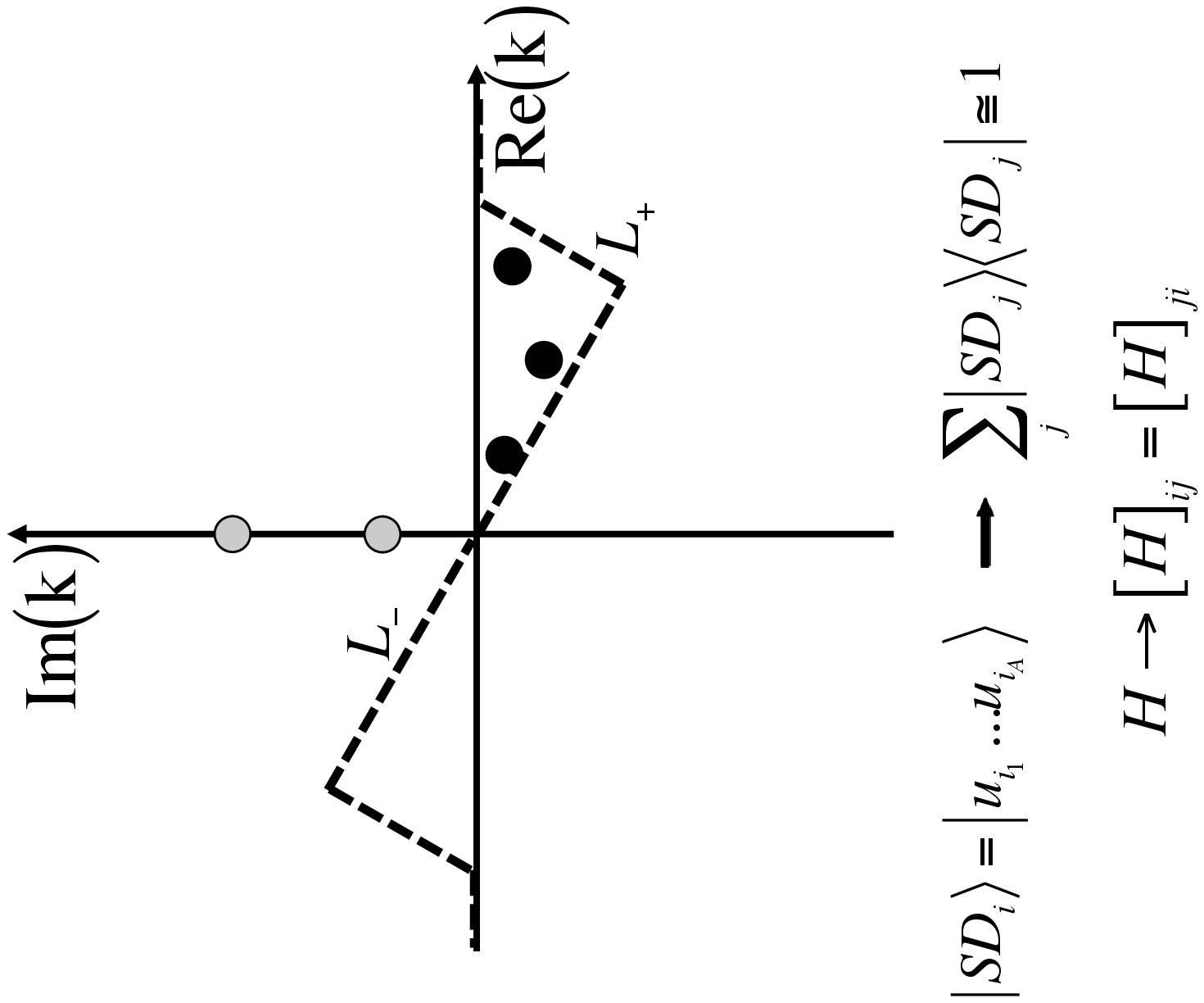}
 \vskip-1cm
\caption{The schematic illustration of a complete s.p. basis in $k$-plane for GSM calculations. The complete many-body basis is spanned by Slater determinants involving nucleons on bound, resonant and continuum s.p. states along the contour $L_+$ in the complex $k$-plane. GSM Hamiltonian matrix is complex symmetric and yields complex energy eigenvalues above the first particle emission threshold.}
\label{fig:1}       
\end{center}
\end{figure*}
Resonance phenomena are generic intrinsic properties of many mesoscopic systems.  What is specific to atomic nuclei are strong nucleon-nucleon correlations which impose a simultaneous description of the configuration mixing and the coupling to decay channels. The fundamental difficulty in the CSM formulation and the reason why the Fano's program\cite{fano} had to wait forty years to find a comprehensive solution is the fact that resonances do not belong to the Hilbert space. Therefore, whatever strategy is chosen to formulate a configuration interaction approach in open quantum systems, a key points are always: (i) the treatment of s.p. resonances in a many-body framework, and (ii) the definition of complete s.p. and many-body bases. 

For the GSM, a s.p. basis is given by the Berggren ensemble\cite{berggren} which consists of Gamow (resonant) states and the non-resonant continuum (see Fig. \ref{fig:1}). (For a detailed description of the GSM see Ref. \cite{Mic09}.) 
The GSM Hamiltonian is Hermitian. However, since the s.p. vectors  have either outgoing or scattering  asymptotics,  the Hamiltonian matrix in GSM is complex symmetric and its eigenvalues are complex above the first particle emission threshold. Hence, both real-energy and complex-energy  CSM formulations lead to a non-Hermitian eigenvalue problem above the  threshold and contain all salient features of an interplay between opposite effects of Hermitian and anti-Hermitian couplings.

\subsection{The Hamiltonian}
The translationally invariant GSM Hamiltonian in intrinsic nucleon-core coordinates of the cluster-orbital shell 
model\cite{Ikeda}, can be written as:
\begin{equation}
H= \sum_{i=1}^{A_{val}}\left[ \frac{p_{i}^{2}}{2\mu_i} + U_{i} \right] + \sum_{i<j}^{A_{val}} \left[ V_{ij} 
+ \frac{1}{M_c} \bf{p}_{i}\bf{p}_{j} 
\right],
\label{GSM_Hamiltonian}
\end{equation}
where $M_c$ is the mass of the core, $\mu_i$ is the reduced mass of either the proton or neutron ($1 / \mu_i =1/m_i + 1/M_{c}$),
$U$ is  the s.p. potential describing the field of the core, $V$ is the two-body residual interaction between
valence nucleons. The last term in (\ref{GSM_Hamiltonian}) represents the recoil term. 

The particle-core interaction is a sum of nuclear and Coulomb terms: $U = U^N + U^C$. The nuclear potential $U^N$ is approximated either by a Woods-Saxon (WS) field with a spin-orbit term\cite{Mic03} or by the 
Gamow-Hartree-Fock (GHF) potential\cite{Mic04}. The Coulomb field $U^C$ is generated by a Gaussian density of $Z_c$ core protons\cite{Mic10}. Similarly, the residual interaction can split into nuclear and Coulomb parts: $V = V^N + V^C$, where $V^N$  is the modified surface Gaussian interaction (MSGI)\cite{Mic10} and $V^C$ is the two-body  Coulomb interaction. 

$V^C$ can be rewritten as: $U^C_{Z_{val}-1} + \left[V^C -  U^C_{Z_{val}-1}\right]^{HO}$,  
where $U^C_{Z_{val}-1}$ takes care of the asymptotic behavior of the Coulomb interaction. 
The second term in this equation and the two-body  recoil term can be expanded in the harmonic oscillator basis\cite{PRC_real_inter_2006,Mic10} which provides  an accurate treatment of the long-range physics of the Coulomb potential. In this work, we took 9 harmonic oscillator shells with the oscillator length $b = 2$\,fm.

\section{N-body GSM reaction wave functions}
The CC framework is a convenient to formulate the GSM description of  reactions involving one proton (neutron) scattering processes. The CC equations are obtained from the following $A$-body matrix elements:
\begin{eqnarray}
\langle \mathcal{A} \{ \langle \Psi_f |^{J_f} \otimes \langle r_f~\ell_f~j_f~\tau_f | \}^{J_A}_{M_A}| H   | \mathcal{A} \{ | \Psi_i \rangle^{J_i} \otimes |r_i~\ell_i~j_i~\tau_i \rangle \}^{J_A}_{M_A}\rangle \label{cc_eqs_NBME} ~ \ ,
\end{eqnarray}
where $|\Psi_i\rangle^{J_i}_{M_i}$~~($|\Psi_f\rangle^{J_f}_{M_f}$) are initial (final) GSM eigenvectors of $(A-1)$-body system, $r_i$, $r_f$ are radial coordinates, 
$\tau_i$, $\tau_f$ are isospin quantum numbers (proton or neutron), and $\ell_i$, $j_i$, $\ell_f$, $j_f$ are angular quantum numbers. All $A$-body wave functions are fully antisymmetrized, as emphasized by the $\mathcal{A}$ symbol.
 
In order to express the antisymmetry in a convenient way, the $|r \ell j \tau \rangle$ channel is expanded in a  s.p. basis of GSM wave functions $u_n(r)$ generated by the s.p. potential $U_{basis}$. This implies for the associated creation operator:
\begin{eqnarray}
a^{\dag}_{r \ell j \tau}=\sum_n u_n(r) ~ a^{\dag}_{n \ell j \tau} \label{a_dagger} ~ \ .
\end{eqnarray}
Hence, the expression of considered $A$-body wave function becomes:
\begin{eqnarray}
\mathcal{A} | \{ | \Psi \rangle^{J} \otimes |r \ell j \tau \rangle \}^{J_A}_{M_A} \rangle 
&=& \sum_n u_n(r) \{a^{\dag}_{n \ell j \tau} | \Psi \rangle^{J} \}^{J_A}_{M_A} ~ \ .
\label{coupled_Psi_rljtau}
\end{eqnarray}

States of the target $(A-1)$-nucleus are eigenstates of the Hamiltonian and can be expressed in the basis of Slater determinant. It is then convenient to express channel states in $(A)$-nucleus in the same Slater determinant basis. Consequently, all formal operations involving many-body operators and channel/target states become straightforward using the second quantization.

\subsection{GSM-CC equations in the coordinate space} \label{cc_eqs_derivation}
In order to evaluate Eq. (\ref{cc_eqs_NBME}), we separate the Hamiltonian $H=T+U_{core}+V_{res}$ into basis and residual parts:
\begin{eqnarray}
H=T+U_{basis} + (V_{res}-U_0)  \label{H_Ubasis_Vres} ~ \ , 
\end{eqnarray}
where $U_{basis}$ is the optimal potential of $A$-particle system and $U_0 = U_{basis} - U_{core}$. $U_{core}$ is the potential generated by the core. The advantage of this decomposition is that $V_{res}-U_0$ is finite-range and $T+U_{basis}$ is diagonal in the basis of Slater determinants used.

Let us consider first: $|\ell_i~j_i~\tau_i \rangle = |\ell_f~j_f~\tau_f \rangle = |\ell j \tau \rangle$ in Eq. (\ref{cc_eqs_NBME}), as $H$ in this case contains infinite-range components leading to Dirac delta's which have to be calculated analytically. In order to derive these expressions, we suppose that only a finite number of Slater determinants appear in target many-body states. This assumption is always valid because GSM eigenvectors expansion coefficients decrease exponentially with the energy of basis scattering Slater determinants and, therefore, can be approximated with any arbitrary precision by a finite expansion of Slater determinants. As a consequence, the antisymmetry between $|u_n \ell j \tau \rangle$ and $| \Psi \rangle^J$ in Eq. (\ref{coupled_Psi_rljtau}) no longer plays any role  for $n$ larger than a given $n_{max}$. This implies that the creation operators in Eq. (\ref{coupled_Psi_rljtau}) can be replaced by the tensor products.

It is convenient to have matrix elements $\langle \alpha \beta | V_{res}-U_0 | \gamma \delta \rangle$ vanished when $n_\alpha > n_{max}$ (same for $\beta, \gamma$ or $\delta$). This is always the case in GSM calculations as one uses a finite model space. It is thus convenient to rewrite the Hamiltonian $H$ of Eq. (\ref{H_Ubasis_Vres}) introducing an operator acting only on the target:
\begin{eqnarray}
H &=& T + U_{basis} + (V_{res}-U_0)^{A-1} + [(V_{res}-U_0) - (V_{res}-U_0)^{A-1}] \ , 
\label{H_Ubasis_Vres_A_minus_one}
\end{eqnarray}
where one defines $(V_{res}-U_0)^{A-1}$ by its action on the non-antisymmetrized $A$-body states $| \Psi \rangle^{J}_{M} \otimes |u_n \ell j m \tau \rangle$:
\begin{eqnarray}
(V_{res}-U_0)^{A-1} (| \Psi \rangle^{J}_{M} \otimes |u_n \ell j m \tau \rangle) = [(V_{res}-U_0) | \Psi \rangle^{J}_{M}] \otimes |u_n \ell j m \tau \rangle \label{Vres_restricted}\ ,
\end{eqnarray}
{\em i.e.}, $(V_{res}-U_0)^{A-1}$ is the finite-range part of $V_{res}-U_0$ acting on $(A-1)$-body states only. Inserting (\ref{H_Ubasis_Vres_A_minus_one}), (\ref{Vres_restricted}) in (\ref{coupled_Psi_rljtau}), one can show that  it is only the sum involving $T + U_{basis} + (V_{res}-U_0)^{A-1}$ and $n \geq 0$ which generates Dirac delta's in the matrix element 
(\ref{cc_eqs_NBME}). Consequently, one can rewrite the matrix element as a sum of two terms, one which is finite and the other which is infinite:
\begin{eqnarray}
&&\langle \mathcal{A} \{ \langle \Psi_f |^{J_f} \otimes \langle r_f \ell j \tau | \}^{J_A}_{M_A}| H | \mathcal{A} \{ | \Psi_i \rangle^{J_i} \otimes |r_i \ell j \tau \rangle \}^{J_A}_{M_A}\rangle  \nonumber \\
&&= \left[ \sum_{n_i n_f}^{n_{max}} u_{n_i}(r_i) ~ u_{n_f}(r_f)  ~
\langle \mathcal{A} \{ \langle \Psi_f |^{J_f} \otimes \langle u_{n_f} \ell j \tau | \}^{J_A}_{M_A}| H | \mathcal{A} \{ | \Psi_i \rangle^{J_i} \otimes | u_{n_i} \ell j \tau \rangle \}^{J_A}_{M_A}\rangle \right. \nonumber \\
&&-\left. \sum_{n \leq n_{max}} u_{n}(r_i) ~ u_{n}(r_f)  ~
\langle \{ \langle \Psi_f |^{J_f} \otimes \langle u_{n} \ell j \tau | \}^{J_A}_{M_A}| T + U_{basis} + (V_{res}-U_0)^{A-1}  | \{ | \Psi_i \rangle^{J_i} \otimes | u_{n} \ell j \tau \rangle \}^{J_A}_{M_A}\rangle \right]
\nonumber \\
&&+ \sum_{n \geq 0} u_{n}(r_i) ~ u_{n}(r_f)  ~
\langle \{ \langle \Psi_f |^{J_f} \otimes \langle u_{n} \ell j \tau | \}^{J_A}_{M_A}| T + U_{basis} + (V_{res}-U_0)^{A-1}  | \{ | \Psi_i \rangle^{J_i} \otimes | u_{n} \ell j \tau \rangle \}^{J_A}_{M_A}\rangle ~ \ . \nonumber \\
\label{H_concise_decomposition} 
\end{eqnarray}

The term in between square bracket is finite and can be calculated using Slater determinant expansion of considered many-body states and employing standard SM formulas. As $| \Psi_i \rangle^{J_i}_{M_i}$ and $| \Psi_f \rangle^{J_f}_{M_f}$ are eigenvectors of $H$, the second term of Eq. (\ref{H_concise_decomposition}) does not vanish if: $$| \Psi_i \rangle^{J_i}_{M_i} = | \Psi_f \rangle^{J_f}_{M_f} = | \Psi \rangle^{J}_{M} \ .$$

If $|\ell_i~j_i~\tau_i \rangle \neq |\ell_f~j_f~\tau_f \rangle$, one can show using the orthogonality of $|\ell_i~j_i~\tau_i \rangle$ and $|\ell_f~j_f~\tau_f \rangle$ that only the sum over $n_i, n_f$ in  Eq. (\ref{H_concise_decomposition})  is nonzero and can be calculated straightforwardly from the Slater determinant expansions of $| \Psi_i \rangle^{J_i}_{M_i}$ and $| \Psi_f \rangle^{J_f}_{M_f}$.

\section{Resolution of the CC equations}
Let us consider the scattering $A$-body state decomposed in reaction channels:
\begin{eqnarray}
| \Phi \rangle = \sum_{c} \int_0^{+\infty} \mathcal{A} | \{ | \Psi_{c} \rangle^{J_{c}} \otimes |r~\ell_{c} j_{c} \tau_{c} \rangle \}^{J_A}_{M_A}\rangle u_{c}(r) r^2 ~ dr ~ \ ,
\end{eqnarray}
where $c$ is  the reaction channel defined by the $(A-1)$-body state $| \Psi_{c} \rangle^{J_{c}}_{M_{c}}$ and the one-body quantum numbers $(\ell_{c},j_{c},\tau_{c})$. $u_{c}(r)$ is the radial amplitude of the $c$ channel to be determined. The CC equations in this basis follow from the Schr{\"o}dinger equation $H | \Phi \rangle = E | \Phi \rangle$:
\begin{equation}
\sum_{c'} \int_0^{+\infty} \langle \mathcal{A} \{ \langle \Psi_c |^{J_c} \otimes \langle r~\ell_c j_c \tau | \}^{J_A} | H - E| \mathcal{A} \{ | \Psi_{c'} \rangle^{J_{c'}} \otimes |r' \ell_{c'} j_{c'} \tau_{c'} \rangle \}^{J_A} \rangle 
                          u_{c'}(r') (r')^2 ~ dr' = 0 \label{cc_eqs_formal_radial} ~ \ . 
\end{equation}

Eq. (\ref{cc_eqs_formal_radial}) is a generalized eigenvalue problem. Indeed, different channels are mutually non-orthogonal because target and projectile are antisymmetrized. 
In order to deal with a standard eigenvalue problem, one introduces the channel overlap matrix:
\begin{equation}
O(n,c,n',c') =  \langle \mathcal{A} \{ \langle \Psi_c |^{J_c} \otimes \langle n~\ell_c j_c \tau | \}^{J_A} | \mathcal{A} \{ | \Psi_{c'} \rangle^{J_{c'}} \otimes |n' \ell_{c'} j_{c'} \tau_{c'} \rangle \}^{J_A} \rangle  \label{overlap_matrix} ~ \ . 
\end{equation}
Eq. (\ref{cc_eqs_formal_radial}) can then be written in a matrix form: $H U = E O U$, where $U = \{ u_c(r) \}_c$ is the vector of considered channels. Introducing: $W = O^{\frac{1}{2}} U$, and the modified Hamiltonian: $H_m = O^{-\frac{1}{2}} H O^{-\frac{1}{2}}$, one obtains the standard eigenvalue problem: $H_m W = E W$.

The overlap matrix $O$ is defined in the Berggren basis and can be calculated using the Slater determinant expansion of channels. As the antisymmetry acts locally, it is convenient to introduce the harmonic oscillator expansion for the finite-range part of $O^{-\frac{1}{2}}$. For this, $O$ is expanded in the basis of harmonic oscillator channels to obtain the finite-range part of $O^{-\frac{1}{2}}$: $\Delta = O^{-\frac{1}{2}} - I_d$.
Then, $H_m$ can be separated into long- and short-range parts: $$H_m = H + H \Delta + \Delta H + \Delta H \Delta ~ ,$$
where all terms involving $\Delta$ are expanded in the harmonic oscillator basis. Thus, the added part of $H$ in $H_m$ can be treated similarly to the short-range residual interaction.

Using results of Sec. (\ref{cc_eqs_derivation}) and the transformation described above, one can write Eq. (\ref{cc_eqs_formal_radial}) as a system of non-local differential equation with respectively local ($V^{(loc)}_{c c'}(r)$) and non-local ($V^{(non-loc)}_{c c'}(r,r') ~ \forall c$) optical potentials:
\begin{eqnarray}
&&\frac{\hbar^2}{2 \mu} \left( -w_c''(r) + \frac{\ell_c(\ell_c+1)}{r^2} w_c(r) \right) + \sum_{c'} \left( V^{(loc)}_{c c'}(r) w_{c'}(r) + \int_0^{+\infty} \!\!\!\! V^{(non-loc)}_{c c'}(r,r') w_{c'}(r') ~ dr' \right) = \nonumber \\
&&=(E - E_{T_c}) ~ w_c(r) 
\label{cc_eqs_diff_radial} ~ \ ,
\end{eqnarray}
where $\mu$ is the reduced mass of the particle and $E_{T_c}$ is the energy of $| \Psi_{c} \rangle^{J_{c}}_{M_{c}}$. 
This set of non-local differential equations is then solved numerically using the modified equivalent potential method which yields equations local without singularities in the potentials\cite{Mic09x}. The initial vector $U = \{ u_c(r) \}_c$ of channel functions is obtained by multiplying solutions of Eq. (\ref{cc_eqs_diff_radial}): $W = \{ w_c(r) \}_c$, by $ O^{-\frac{1}{2}}$: $U = W + \Delta W$.

\section{Discussion of results for $^6$He+p reaction}
Results presented in this section correspond to $^6$He target nucleus in 2 states: $J^{\pi}=0_1^+$ and $2_1^+$. The configuration space for neutrons correspond to $0p_{3/2}$ resonance and 17 states of a discretized $p_{3/2}$ contour in the complex $k$ plane. The configuration space for protons includes $0p_{1/2}$ and $0p_{3/2}$ resonances and 17 states of a discretized contour for each resonance. Moreover, for protons we include partial waves $s,d,f,g,h$ which are decomposed using a real-energy contour as in this case the resonant poles are high in energy and very broad. 

The s.p. basis in $^6$He and $^7$Li is generated by the same GHF potential, produced by the WS potential of the core and the MSGI two-body interaction between valence nucleons. Parameters of the core potential: the radius $R=1.993$ fm (1.954 fm), the depth of the central part $V_0=50.188$ MeV (49.268 MeV), the diffuseness $d=0.497$ fm (0.657 fm), and the spin-orbit strength $V_{so}=7.488$ MeV (8.23 MeV) for protons (neutrons) have been chosen to fit $p_{1/2}$ and $s_{1/2}$ phase shifts in $^4$He+p and $^4$He+n, and energies/widths of $3/2_1^-$ and $1/2_1^-$ resonances in $^5$He and $^5$Li. The radius of the Coulomb potential in this calculation is $R_C=1.952$ fm. Parameters of the MSGI interaction have been chosen to reproduce energies of the states in $^6$He ($0_1^+$ and $2_1^+$) and  in $^7$Li ($3/2_1^-$, $1/2_1^-$, $7/2_1^-$ and $5/2_1^-$) relative to the $^4$He core. 
Eq. (\ref{cc_eqs_diff_radial}) are then solved using the multidimensional variant of one-dimensional iterative procedure\cite{Mic09x} where the input functions for the first iteration come from the diagonalization of $H_m$ in the Berggren basis. 

\subsection{Numerical tests}
Before showing results for $^6$He+p scattering, it is instructive to compare the spectra of GSM-CC calculations with those obtained in GSM by a direct diagonalization in Berggren basis. One should remember that the configuration space in both calculations is in general not the same. 
\begin{figure*}
\begin{center}
 \includegraphics[width=11.5cm,angle=00]{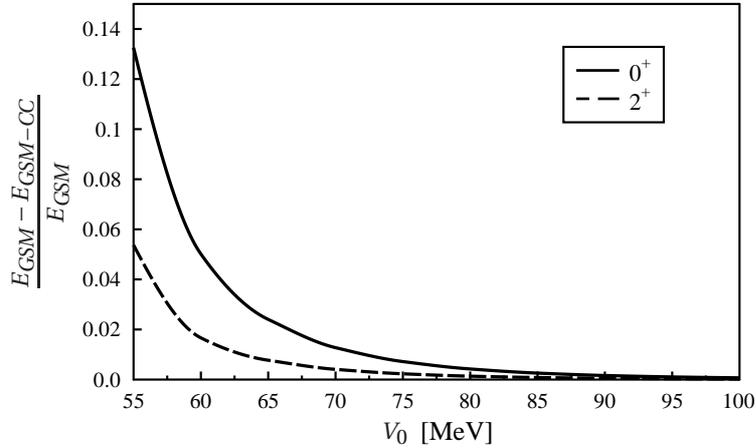}
\vskip-0.5cm
\caption{Relative energy difference between GSM and GSM-CC calculations for $^6$He in $J^{\pi}=0^+$ (the solid line) and $2^+$ (the dashed line) states as a function of the depth of the WS potential generated by $^4$He core. The model space is spanned by $0p_{3/2}$ resonance and 30 $p_{3/2}$ states of the discretized continuum. As a two-body interaction, the MSGI interaction with coupling strengths: $V(J=0,T=1)=-20$MeV$\times$fm$^3$ and $V(J=2,T=1)=-15$MeV$\times$fm$^3$ is used.}
\label{fig:2}       
\end{center}
\end{figure*}
\begin{figure*}
\begin{center}
 \includegraphics[width=11.5cm,angle=00]{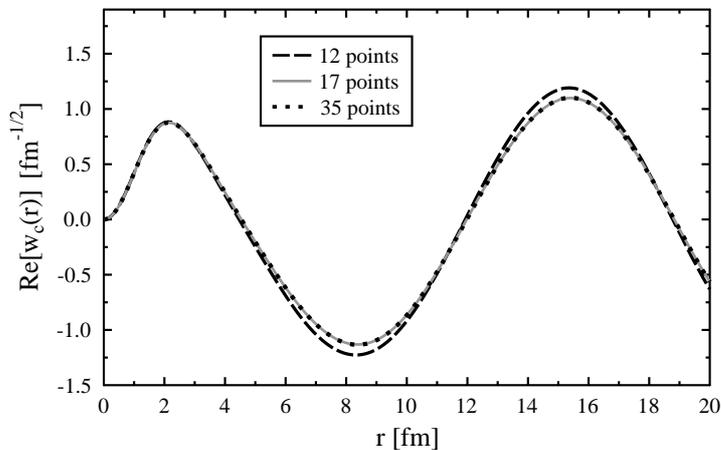}
 \vskip-0.5cm
\caption{The real part of $3/2^-$ scattering wave function at the CM energy $E_{CM}$=5 MeV in the entrance channel $[ ^6{\rm He}(0_1^+) \otimes \pi p_{3/2} ]$ is plotted for different discretization densities of the $p_{3/2}$ contour in the complex $k$-plane.}
\label{fig:3}       
\end{center}
\end{figure*}
Fig. \ref{fig:2} compares results of the GSM calculation for $^6$He with the GSM-CC results for a system $^5$He$(3/2_1^-)$+n in a model space including $0p_{3/2}$ s.p. resonance and $p_{3/2}$ continuum discretized with 30 points. One can see that if the continuum coupling  is weak, {\em i.e.} for deep WS potentials, then the relative difference between GSM and GSM-CC calculations tends to zero. On the contrary, for weakly bound systems the GSM-CC approach with a limited number of reaction channels can be a poor approximation of the GSM calculation in the complete many-body space. 
 
The discretization density of the Berggren basis is an essential ingredient of the GSM-CC calculations. 
Fig. \ref{fig:3} shows a convergence of the $3/2^-$ scattering wave function in $^7$Li as a function of the number of $p_{3/2}$ states on the discretized contour. One can see that the scattering wave function is fully converged with 17 continuum states. 

\subsection{Elastic and inelastic $^6$He+p cross section}
\begin{figure*}
 \includegraphics[width=16.5cm,angle=00]{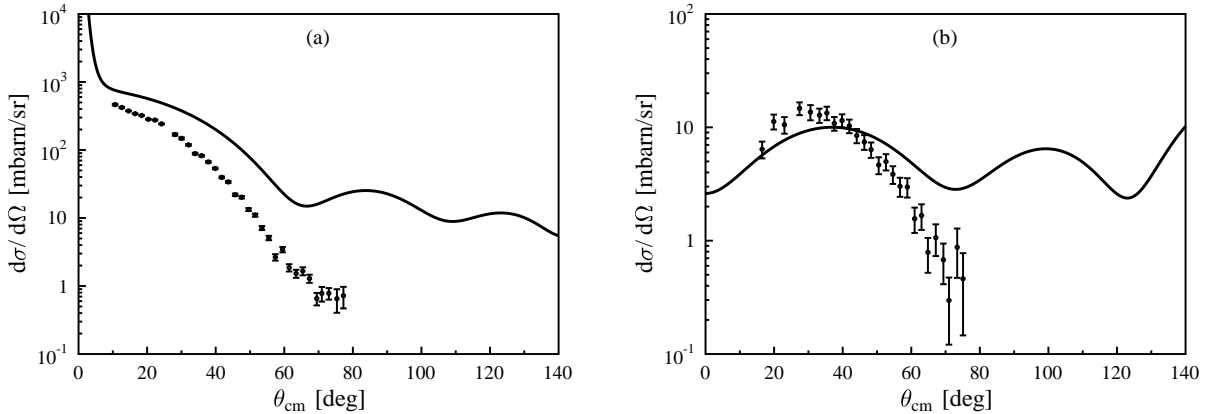}
 \vskip-0.5cm
\caption{Elastic (left panel) and inelastic (right panel) GSM-CC angular cross sections for the reaction $^6$He+p at the CM energy $E_{CM}$=35.1 MeV. The experimental data are taken from Ref. \cite{Lag01}.}
\label{fig:4}       
\end{figure*}
Angular cross-sections calculated in GSM-CC for $^6$He+p reaction at $E_{CM}=35.1$ MeV are in the qualitative agreement with experimental data\cite{Lag01} (see Fig. \ref{fig:4}). Quantitative discrepancies between theory and experiment may have several origins.
Firstly, the excitation energy in Fig. \ref{fig:4} is such that  the contribution of $^4$He core excitations in this reaction cannot be neglected. Secondly, the chosen effective interaction is rather schematic.
Thirdly and most importantly, the target is a weakly bound halo state whereas the reaction product ($^7$Li) in its ground state is well bound with strong $^3$H cluster correlations due to the proximity of $^3$H decay threshold. This implies that the GSM-CC configuration space generated by adding one proton to $^6$He in 2 discrete states ($0_1^+$ ground state and $2_1^+$ resonance) only,  may by insufficient to produce configurations obtained by a direct diagonalization of $^7$Li in the Berggren basis. Indeed, the configuration spaces in GSM-CC and GSM become equivalent if {\em all} discrete and continuum states of $^6$He are used to generate $^7$Li configurations. Including only few states of $^6$He in GSM-CC approach requires a strong renormalization of the effective nucleon-nucleon interaction (the optical potentials) to compensate for the missing configurations. By construction, such a renormalization will generate identical energy spectra for both full space and restricted space GSM-CC calculations. However, the reaction cross-sections in these two settings will be different. 
\begin{table}
\caption{\label{tab:1}GSM and GSM-CC spectra for low-lying states of $^7$Li are compared with the experimental data. For the details of these calculations, see the discussion in the text.}
\vskip 0.5cm
\begin{center}
\begin{tabular}{llll}
\br
$J^{\pi}$ & $E_{GSM}$ (MeV) & $E_{GSM-CC}$ (MeV) & $E_{exp}$ (MeV)\\
\mr
$3/2^-$ & -17.83 & -10.946 & -10.949 \\
$1/2^-$ & -21.18 & -10.469 & -10.471 \\
$7/2^-$ & -21.01 & -6.307 & -6.297 \\
$5/2^-$ & -31.26 & -4.509 & -4.345 \\
\br
\end{tabular}
\end{center}
\end{table}
The comparison between energy spectra obtained for the same Hamiltonian either by a direct diagonalization in Berggren basis or by solving GSM-CC equations for a limited number of many-body target states gives indication of how strong should be the renormalization of effective two-body interaction (the optical potentials) due to the neglected channels.  In Table \ref{tab:1}, we compare energy spectra obtained in GSM and GSM-CC approaches using the same MSGI interaction which was fitted in the GSM-CC approach to reproduce the low-lying states of $^6$He and $^7$Li. One may see that  $^7$Li configurations involving $^6$He scattering states which are missing in the GSM-CC calculation, lead to a dramatic lowering of the GSM eigenenergies and, hence, are essential to understand the dynamics in the $^6$He+p reaction. 

Results shown in Table \ref{tab:1} demonstrate that none of the CC approaches which neglects continuum states of the target can provide a realistic description of the proton scattering on weakly-bound neutron-rich nuclei. Similar conclusion about the importance of high-lying states in the continuum of a target $(A-1)$-nucleus on low-lying properties of the $(A)$-nucleus have been made recently in the Shell Model Embedded in the Continuum\cite{Oko08}.

\section{Conclusions}
We have proposed the unified description of nuclear structure and nuclear reactions in the framework of GSM which includes continuum couplings in the many-body framework. This CC formulation of the GSM can be employed further in the No-Core GSM framework\cite{papa12} with the bare interaction between free nucleons. The great advantage of the GSM-CC formalism is that the approximation of neglecting high-lying target states  can be checked by comparing the calculated GSM-CC energy spectrum with those obtained in (complete) GSM calculation. In this way, one can quantify the role of these neglected configurations in the $(A)$-particle wave function and estimate the scale of renormalization corrections in the microscopic optical potentials. 

In spite of recent developments in {\it ab initio} description of nuclear states and progress in open quantum system formulation of the nuclear many-body problem, the microscopic description of nuclear reactions with weakly bound targets continues to be a formidable challenge as it requires including large number of target states to obtain convergent results. In this respect, the task of a unified microscopic description of nuclear structure and nuclear reactions with weakly bound exotic nuclei remains still a distant perspective.

\ack
This paper is written to honor important and numerous contributions of Jerry P. Draayer to the nuclear many-body theory.
This work has been supported in part by the by the MNiSW grant No. N N202 033837; the Collaboration COPIN-GANIL; and  U.S. Department of Energy under Contract Nos. DE-FG02-96ER40963 (University of Tennessee) and DE-FG02-10ER41700 (French-U.S. Theory Institute for Physics with Exotic Nuclei).

\end{document}